%Paper: hep-th/9503016
%From: Dan Kabat <kabat@physics.rutgers.edu>
%Date: Fri, 3 Mar 1995 00:28:45 -0500
%Date (revised): Wed, 8 Mar 1995 16:42:20 -0500
%Date (revised): Thu, 20 Jul 1995 15:38:56 -0400

\input harvmac
\input epsf
\def\epsfsize#1#2{\ifnum #1 > \hsize \hsize \else #1 \fi}
\newcount \pageit
\footline={\tenrm\hss \ifnum\pageit=0 \hfill \else \number\pageno \fi\hss}
\pageit=0
\pageno=0
\centerline{\bf BLACK HOLE ENTROPY AND ENTROPY OF ENTANGLEMENT}
\vskip 24pt
\centerline{Daniel Kabat}
\vskip 12pt
\centerline{\it Department of Physics and Astronomy}
\centerline{\it Rutgers University}
\centerline{\it Piscataway, NJ 08855--0849}
\vskip 12pt
\centerline{\it kabat@physics.rutgers.edu}
\vskip 0.9 true in
\centerline{\bf Abstract}
\bigskip
\leftskip = 0.5 true in
\rightskip = 0.5 true in
\noindent
We compare the one-loop corrections to the entropy of a black hole,
from quantum fields of spin zero, one-half, and one, to the entropy of
entanglement of the fields.  For fields of spin zero and one-half the
black hole entropy is identical to the entropy of entanglement.  For
spin one the two entropies differ by a contact interaction with the
horizon which appears in the black hole entropy but not in the entropy
of entanglement.  The contact interaction can be expressed as a path
integral over particle paths which begin and end on the horizon; it is
the field theory limit of the interaction proposed by Susskind and
Uglum which couples a closed string to an open string stranded on the
horizon.
\leftskip = 0.0 true in
\rightskip = 0.0 true in
\vfill
\vskip -28 pt
\noindent hep-th/9503016 \hfill March 1995
\smallskip
\noindent RU--95--06
\eject
\pageit=1

\def\half{{1 \over 2}}

\def\laplace{{\kern1pt\vbox{\hrule height 1.2pt\hbox{\vrule width 1.2pt\hskip
  3pt\vbox{\vskip 6pt}\hskip 3pt\vrule width 0.6pt}\hrule height 0.6pt}
  \kern1pt}}
\def\scriptlap{{\kern1pt\vbox{\hrule height 0.8pt\hbox{\vrule width 0.8pt
  \hskip2pt\vbox{\vskip 4pt}\hskip 2pt\vrule width 0.4pt}\hrule height 0.4pt}
  \kern1pt}}
\def\slash#1{{\rlap{$#1$}\thinspace /}}

\def\real{{{\rm R}\llap{\vrule height 6.9pt width 0.6pt depth -.3pt
  \phantom t}}}
\def\integer{{\rlap{\rm Z} \hskip 1.6pt {\rm Z}}}

\def\nullcases#1{\left.\,\vcenter{\normalbaselines\mathsurround=0pt
    \ialign{$##\hfil$&\quad##\hfil\crcr#1\crcr}}\right.}

\font\eightrm=cmr8

\def\kp{{{\bf k}_\perp}}
\def\xp{{{\bf x}_\perp}}

\def\pr{{\it Phys.~Rev.~}}
\def\prl{{\it Phys.~Rev.~Lett.~}}

\def\np{{\it Nucl.~Phys.~}}
\def\pl{{\it Phys.~Lett.~}}
\def\prep{{\it Phys.~Rep.~}}
\def\cmp{{\it Commun.~Math.~Phys.~}}

\def\cqg{{\it Class.~Quant.~Grav.~}}

\def\ap{{\it Ann.~Phys.~}}

\nref\BTZ{M.~Ba\~nados, C.~Teitelboim, and J.~Zanelli, \prl {\bf 72}, 957
(1994), gr-qc/9309026.}
\nref\CT{S.~Carlip and C.~Teitelboim, gr-qc/9312002.}
\nref\Israel{W.~Israel, \pl {\bf A57}, 107 (1976).}
\nref\tH{G.~'t Hooft, \np {\bf B256}, 727 (1985).}
\nref\BKLS{L.~Bombelli, R.~K.~Koul, J.~Lee, R.~D.~Sorkin, \pr {\bf D34},
 373 (1986).}
\nref\Sred{M.~Srednicki, \prl {\bf 71}, 666 (1993).}
\nref\SussUg{L.~Susskind, hep-th/9309145\semi  L.~Susskind and J.~Uglum,
\pr {\bf D50}, 2700 (1994), hep-th/9401070.}
\nref\CalWil{C.~Callan and F.~Wilczek, \pl {\bf B333}, 55 (1994),
hep-th/9401072.}
\nref\KabStr{D.~Kabat and M.~Strassler, \pl {\bf B329}, 46 (1994),
hep-th/9401125.}
\nref\DowEnt{J.~S.~Dowker, \cqg {\bf 11}, L55 (1994), hep-th/9401159.}
\nref\renorm{S.~Solodukhin, \pr {\bf D51}, 609 (1995), hep-th/9407001\semi
D.~Fursaev, hep-th/9408066\semi
J.~G.~Demers, R.~Lafrance, and R.~Myers, gr-qc/9503003.}
\nref\BorByt{M.~Bordag and A.~Bytsenko, gr-qc/9412054.}
\lref\BirrDav{N.~D.~Birrell and P.~C.~W.~Davies, {\it Quantum Fields in
 Curved Space} (Cambridge, 1982) section 6.2.}
\lref\Dab{Orbifolds may be used at a discrete infinity of temperatures.
A.~Dabholkar, hep-th/9408098.}
\lref\thermal{See for example G.~L.~Sewell, \ap {\bf 141}, 201 (1982)\semi
S.~Fulling and S.~Ruijsenaars, \prep {\bf 152}, 135 (1987).}
\lref\Unruh{W.~G.~Unruh, \pr {\bf D14}, 870 (1976).}
\lref\LarWil{F.~Larsen and F.~Wilczek, hep-th/9408089.}
\lref\OnCones{G.~Cognola, K.~Kirsten, and L.~Vanzo, \pr {\bf D49},
1029 (1994), hep-th/9308106\semi
D.~Fursaev, \cqg {\bf 11}, 1431 (1994), hep-th/9309050.}
\lref\Dowker{J.~Dowker, {\it J.~Phys.} {\bf A 10}, 115 (1977)\semi
J.~Dowker, \pr {\bf D36}, 3742 (1987).}
\lref\DesJac{S.~Deser and R.~Jackiw, \cmp {\bf 118}, 495 (1988).}
\lref\KayStu{B.~S.~Kay and U.~M.~Studer, \cmp {\bf 139}, 103 (1991).}
\lref\Ramond{P.~Ramond, {\it Field Theory: A Modern Primer} (Addison-Wesley,
1989) section 3.7.}
\lref\Barb{J.~L.~F.~Barbon, \pr {\bf D50}, 2712 (1994), hep-th/9402004.}
\lref\Emp{R.~Emparan, hep-th/9407064.}
\lref\DeAlOh{S.~P.~de Alwis and N.~Ohta, hep-th/9412027.}
\lref\BalCE{A.~P.~Balachandran, L.~Chandar, and E.~Ercolessi,
hep-th/9411164.}
\lref\Vass{Gauge fields on non-trivial space-times have been discussed
recently by D.~V.~Vassilevich, gr-qc/9404052; D.~V.~Vassilevich,
gr-qc/9411036.}
\lref\Poly{A.~M.~Polyakov, {\it Gauge Fields and Strings} (Harwood 1987),
sections 9.1 -- 9.3.}
\lref\HoLaWi{C.~Holzhey, F.~Larsen, and F.~Wilczek, \np {\bf B424}, 443
(1994), hep-th/9403108.}
\lref\FermiSchr{R.~Floreanini and R.~Jackiw, \pr {\bf D37}, 2206 (1988)\semi
R.~Jackiw, in {\it Field Theory and Particle Physics},
O.~Eboli, M.~Gomez, and A.~Santoro, eds. (World Scientific, 1990).}
\lref\GerJac{P.~de Sousa Gerbert and R.~Jackiw, \cmp {\bf 124}, 229 (1989).}
\lref\KSS{D.~Kabat, S.~Shenker, and M.~Strassler, hep-th/9506182.}

\newsec{Introduction and Overview}

Black holes have thermal properties.  This remarkable fact can be
established by analogy with an ordinary thermodynamic system, where
temperature is the inverse periodicity in Euclidean time, and entropy
is the variation of free energy with respect to temperature.  The
temperature of a black hole is fixed by requiring that the
Schwarzschild metric yield a smooth solution of the Einstein equations
when continued to imaginary time.  This forces the periodicity of the
Euclidean time coordinate to be the inverse Hawking temperature.  The
entropy of a black hole is then found by varying the periodicity in
Euclidean time while holding the geometry on a spatial slice fixed, a
procedure which introduces a conical singularity at the horizon.  The
classical action for gravity evaluated on such a space leads to the
usual expression for black hole entropy \refs{\BTZ,\CT}, while quantum
corrections to the classical entropy result from fluctuations of the
metric or matter fields in the background with the conical
singularity.  We work in the limit of infinitely massive black holes,
so that curvature vanishes except possibly on the horizon, and the
inverse Hawking temperature becomes the $2 \pi$ periodicity of the
plane in polar coordinates.

Quantum corrections to the entropy from matter fields have been
extensively studied \refs{\Israel{--}\BorByt}.  It has been realized
that, in some cases, these corrections have a state counting
interpretation as entropy of entanglement, or equivalently, as Rindler
thermal entropy.  A quantum field in its Minkowski vacuum state has
correlations between degrees of freedom located on opposite sides of
an imaginary boundary, so measurements made only on one side of the
boundary see a mixed state, with a corresponding entropy of
entanglement.  The density matrix which describes the mixed state is a
thermal density matrix in Rindler coordinates, and entropy of
entanglement may be equivalently understood as the thermal entropy
which the field carries in Rindler space.

In this paper we explore the relationship between black hole entropy
and entropy of entanglement for matter fields of spin zero, one-half,
and one.  We distinguish between black hole entropy, given by the
response of the field to a conical singularity, and entropy of
entanglement, obtained from the density matrix which describes the
vacuum state of the field as observed from one side of a boundary in
Minkowski space.  For spins zero and one-half, there is no need for
the distinction: the one loop correction to the black hole entropy is
equal to the entropy of entanglement.  At spin one, the black hole
entropy is equal to the entropy of entanglement plus a contact
interaction with the horizon.  The contact interaction cannot be
interpreted as entropy of entanglement in quantum field theory; indeed
we will see that it makes the one loop correction to the black hole
entropy negative.

The first hint that something non-trivial happens for spin one comes
from its one-loop renormalization of Newton's constant.  The effective
action from integrating out a matter field in a curved background may
be expanded in derivatives of the metric.
$$\eqalign{
\beta F &= \int d^dx \sqrt{g} {\cal L}_{\rm eff} \cr
{\cal L}_{\rm eff} &= - \half \int_{\epsilon^2}^\infty {ds \over
 (4 \pi s)^{d/2}} e^{-s m^2} \left( {c_0 \over s} + c_1 R +
 {\cal O}(s)\right)\cr}
$$
We must evaluate this effective action on a space which is a cone of
deficit angle $2\pi - \beta$ times a transverse flat
$(d-2)$-dimensional space with area $A_\perp$.  Susskind and Uglum
\SussUg~have argued that for infinitesimal deficit angles only the
Einstein-Hilbert term in the effective action contributes, so that we
only need to know the coefficient $c_1$ \BirrDav.
$$
c_1 = \cases{
   {1 \over 6} & minimally coupled scalar \cr
\noalign{\vskip 4pt}
   {1 \over 12} 2^{\left[{d / 2} \right]} & Dirac fermion (with
         $2^{\left[d/2\right]}$ components) \cr
\noalign{\vskip 4pt}
   {d-2 \over 6} - 1 & abelian gauge field including ghosts \cr}
$$
The integral of the scalar curvature is proportional to the deficit
angle of the cone, $\int d^d x \sqrt{g} R = 2 A_\perp (2 \pi -
\beta)$.  At the on-shell temperature $\beta = 2 \pi$, the contribution
of the field to the entropy of a black hole is given by
\eqn\RenEnt{\eqalign{
S &= \left.\left(\beta {\partial \over \partial \beta} - 1\right)
     \right\vert_{\beta = 2 \pi}(\beta F) \cr
  &= 2 \pi c_1 A_\perp \, \int_{\epsilon^2}^\infty {ds
     \over (4 \pi s)^{d/2}} e^{-s m^2}\,.\cr}}
Note that the contribution to the entropy is proportional to the
horizon area (set $A_\perp = 1$ in $d=2$).  It is ultraviolet
divergent; $\epsilon$ is an ultraviolet cutoff put in by hand.  In
$d=2$ it also diverges in the infrared and the mass $m$ must be kept
non-zero to provide a cutoff.

For $d < 8$ a spin one field makes a negative contribution to the
coefficient of the Einstein-Hilbert term and also, therefore, to the
black hole entropy.  This negative contribution is responsible for the
non-renormalization of Newton's constant in certain supersymmetric
theories.  It means, however, that the contribution to the black hole
entropy from a spin one field cannot be identified solely with entropy
of entanglement or Rindler thermal entropy, both of which are
intrinsically positive quantities.

To understand this better we turn to the work of Susskind and Uglum
\SussUg, who advocate that black hole entropy can be
understood within string theory.  The string diagrams they claim are
responsible for black hole entropy are shown in Fig.~1.  The genus
zero diagram if time sliced with respect to the polar angle around the
horizon describes the propagation of an open string with its ends
stuck on the horizon.  There are two classes of diagrams at one loop.
The first class describes the propagation of a closed string around
the horizon, and has a thermal interpretation as a string at a
position-dependent proper temperature.  The second class describes an
interaction between a closed string and an open string stranded on the
horizon.  It does not have a thermal interpretation, and can only be
viewed as an interaction correction to the entropy which is present at
genus zero.  Unfortunately it is difficult to give a precise meaning
to these string diagrams, since they involve defining string theory in
an off-shell background \Dab.

At low energies, the string diagrams reduce to the particle diagrams
of Fig.~2.  The physical interpretation of the classical entropy as
counting configurations of stranded strings is lost.  At one loop we
have a thermal interpretation of a particle encircling the horizon,
but also expect to find a contact interaction with the horizon.
Susskind and Uglum have suggested that this contact term is
responsible for the non-renormalization of Newton's constant in
certain supersymmetric theories.
\goodbreak
\midinsert
\epsfbox{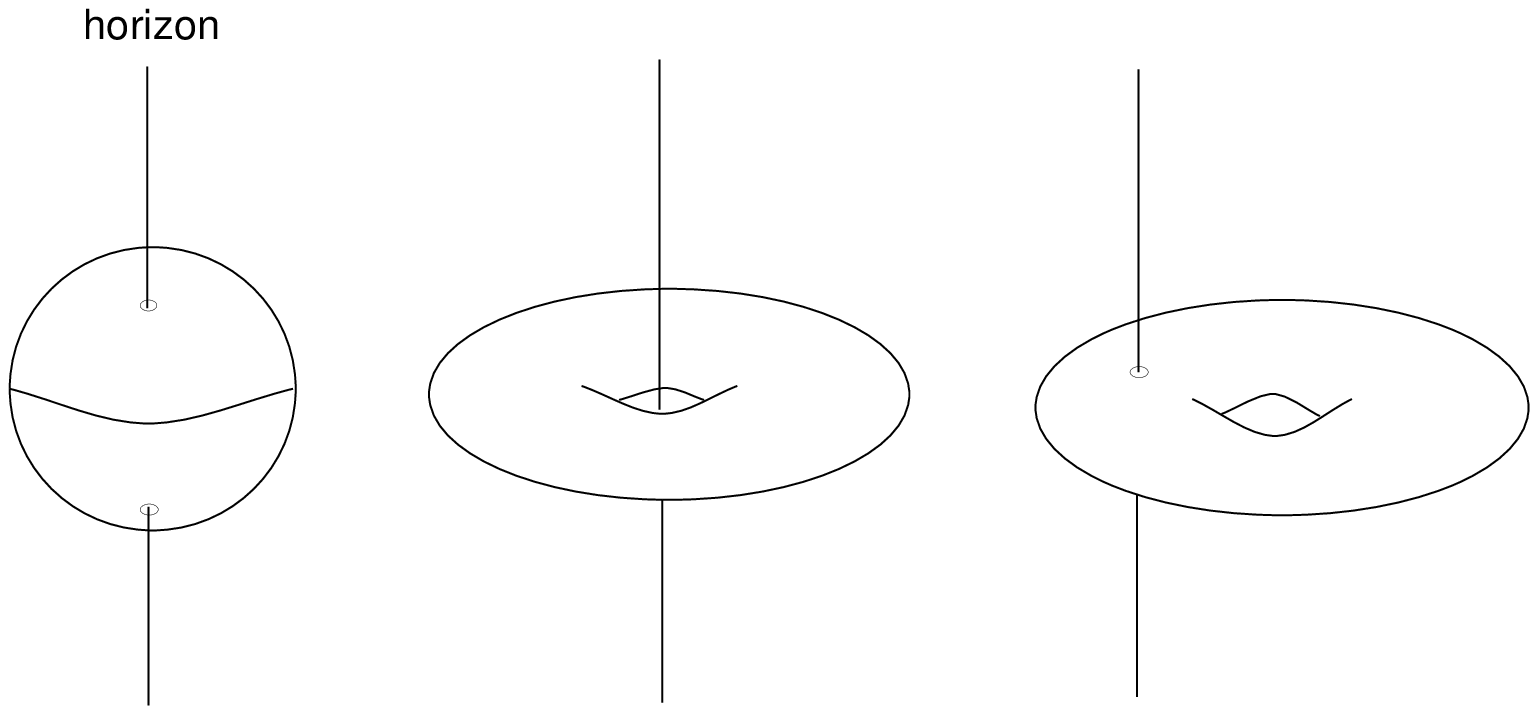}
\centerline{Fig.~1.  String diagrams responsible for the black hole
entropy up to one loop.}
\endinsert
\midinsert
\epsfbox{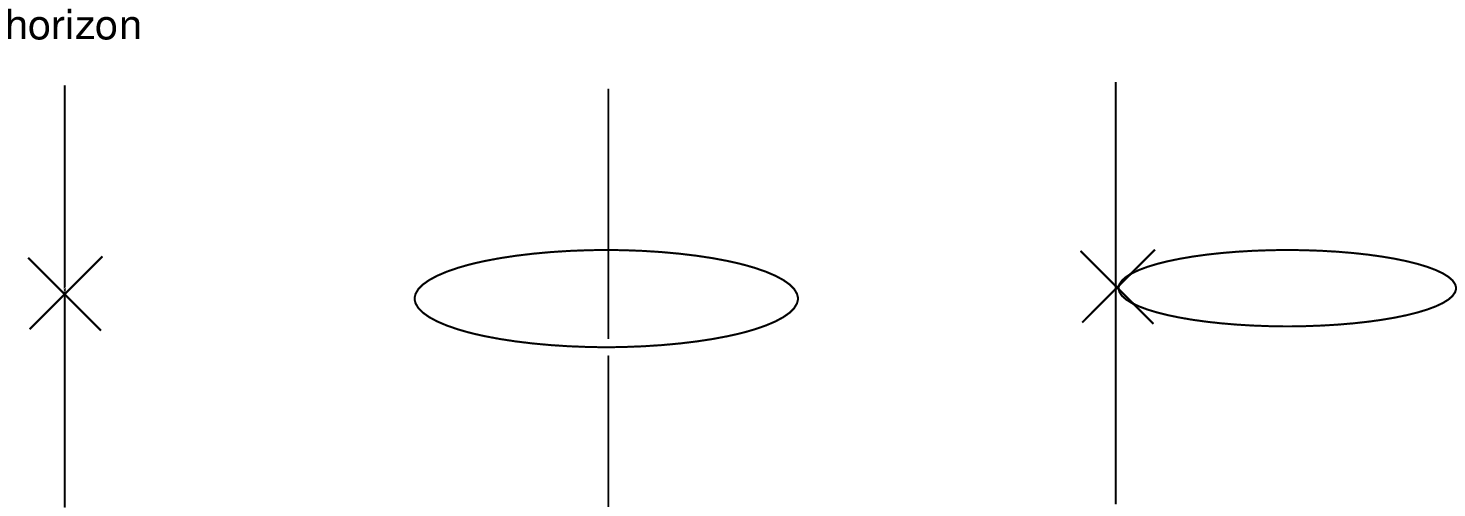}
\centerline{Fig.~2.  Low energy limit of the string diagrams.}
\endinsert

\goodbreak
Motivated by these particle diagrams we calculate the partition
function on a cone for fields of spin zero, one-half, and one.  We now
summarize our main results.  For a scalar field the relevant one
loop determinant may be expressed in terms of a single particle path
integral, which can be explicitly evaluated.
$$\eqalign{
\beta F_{\rm scalar} &= \half \log \det (-\laplace+m^2)\cr
&= - \half \int_{\epsilon^2}^\infty {ds \over s}\int\limits_{\hbox{\eightrm
         closed loops}}
       {\cal D}x(\tau)\,\, \exp - \int_0^s d\tau \left({1\over 4} g_{\mu\nu}
       \dot{x}^\mu \dot{x}^\nu + m^2\right)\cr
&= - {\pi^2 \over 3 \beta} A_\perp \left(1-\left({\beta \over 2 \pi}\right)^2
          \right) \int_{\epsilon^2}^\infty {ds \over (4\pi s)^{d/2}}
          e^{-s m^2}\cr}
$$
In the final line we drop a divergent cosmological constant, which
does not affect the entropy.  In two dimensions $A_\perp = 1$, and the
mass must be kept non-zero as an infrared regulator.  Note that the
on-shell entropy agrees with the result obtained above from the
renormalization of Newton's constant; this shows that the argument of
Susskind and Uglum that only the Einstein-Hilbert term contributes is
valid.  We will show that this entropy is equal to the entropy of
entanglement, as well as the Rindler thermal entropy, of a scalar
field.

For an Abelian gauge field we have the following one loop determinant.
We use a covariant gauge, neglect the cosmological constant, and introduce
an infrared regulating mass which may be set to zero in greater than two
dimensions.
$$\eqalign{
\beta F_{\rm gauge} &= \half \log \det \left(g^{\mu\nu}(-\laplace+m^2)
 - R^{\mu\nu} + \left(1-{1 \over \xi}\right)\nabla^\mu\nabla^\nu\right)
 - \log \det (-\laplace + m^2)\cr
&= (d-2) \beta F_{\rm scalar}\cr
&\quad + \beta\int_{\epsilon^2}^\infty ds \!\!
 \int\limits_{\,\,\,\,x(0) = x(s) \in\, {\rm horizon}}
 \!\!\!\!\!\!\!\!{\cal D}x(\tau)\,\, \exp - \int_0^s d\tau
   ({1\over 4} g_{\mu\nu} \dot{x}^\mu \dot{x}^\nu + m^2)\cr
&\quad- \beta A_\perp \int_{\epsilon^2}^\infty {ds \over (4 \pi s)^{d/2}}
 e^{-s m^2}\cr}
$$
The first term is the free energy of $d-2$ scalar fields,
corresponding to the $d-2$ physical polarizations of a gauge field.
The next term is a surface term at the horizon, expressed as an
integral over particle paths which begin and end on the horizon.  The
last term is actually a surface term at infinity; it will cancel the
the surface term at the horizon when $\beta=2\pi$.  The path integrals
may be explicitly evaluated.
$$
\beta F_{\rm gauge}= (d-2) \beta F_{\rm scalar} + A_\perp (2\pi - \beta)
      \int_{\epsilon^2}^\infty {ds \over (4 \pi s)^{d/2}} e^{-s m^2}
$$
This gives the same on-shell entropy as was obtained above from the
renormalization of Newton's constant.  We will show that the bulk term
in the black hole entropy is equal to the entropy of entanglement of
the field, and can also be understood as bulk thermal entropy in
Rindler space.  The surface term, on the other hand, does not admit a
thermal or state counting interpretation; it is not present in the
entropy of entanglement.

In two dimensions a Dirac fermion has the same partition function on a
cone as a scalar field, $\beta F_{\rm Dirac} = \beta F_{\rm scalar}$
(modulo a cosmological constant).  This gives the same on-shell
entropy as was obtained above from the renormalization of Newton's
constant.  It is equal to the entropy of entanglement, as well as the
Rindler thermal entropy, of the Dirac field.

The rest of this paper develops these claims in detail.  In the next
section we consider scalar fields; many of the results exist in the
literature but we will need them for reference.  The following two
sections treat Abelian gauge fields and Dirac fermions.  In the final
section we summarize our results and speculate about their
implications.

\newsec{Scalar Fields}

We wish to show that the entropy of entanglement of a scalar field is
identical to the entropy which it contributes to a black hole.  We'll
begin from the definition of entropy of entanglement, and show that it
can be calculated from the partition function on a cone.  For
simplicity we work in two dimensions.

First, we must find the Minkowski vacuum state of the field.
The vacuum wavefunctional is given by a Euclidean path integral on the
upper half plane.
$$
\psi[\phi_0(x)] = \int\limits_{\phi(\tau=0,x)=\phi_0(x)} \!\!\!\!\!\!
                   {\cal D} \phi\,
                   \exp - \int_0^\infty d\tau \int_{-\infty}^\infty dx
                   \left(\half g^{\mu\nu}\partial_\mu\phi\partial_\nu\phi
                   + \half m^2 \phi^2\right)
$$
A classical field $\phi_{\rm cl}$ obeying the equations of motion and
the relevant boundary conditions, $\phi_{\rm cl}(\tau=0,x) =
\phi_0(x)$ and $\phi_{\rm cl} \rightarrow 0$ as $\tau \rightarrow
\infty$, may be given in terms of the Dirichlet Green's function on
the upper half plane, which can be constructed by the method of
images.
$$\eqalign{
&\phi_{\rm cl}(\tau,x) = \int_{-\infty}^{\infty} dx'
   \partial_{\tau'}\big\vert_{\tau'=0} G_D(\tau,x\vert\tau',x')
   \phi_0(x')\cr
&G_D(\tau,x\vert\tau',x') = G(\tau,x\vert\tau',x') -
   G(\tau,x\vert-\tau',x')\cr}
$$
The Gaussian path integral is easily evaluated.
$$
\psi[\phi_0(x)] = \det{}^{1/2} G_D  \,\, \exp \int_{-\infty}^\infty dx
      \int_{-\infty}^\infty dx' \phi_0(x) \partial_\tau
      \big\vert_{\tau=0} \partial_{\tau'}
      \big\vert_{\tau'=0} G(\tau,x\vert\tau',x')
      \phi_0(x')
$$

Next we express this vacuum wavefunctional in a more convenient basis.
In polar coordinates on the plane, the quantum mechanical rotation
generator (Rindler Hamiltonian) $H_R = \sqrt{-r\partial_r r\partial_r
+ m^2 r^2}$ is self-adjoint with respect to the inner product
$<\phi_1\vert\phi_2> = \int_0^\infty {dr \over r} \phi_1^* \phi_2$;
its eigenfunctions obey the expected completeness relations.
\eqn\ScalarHR{\eqalign{
&\phi_E(r) = {1 \over \pi} \sqrt{2E \sinh \pi E} \, K_{iE}(mr)\cr
&H_R \phi_E(r) = E \phi_E(r) \qquad 0 < E < \infty \cr
&\int_0^\infty {dr \over r} \phi_E(r) \phi_{E'}(r) = \delta(E-E')\cr}}
The Green's function takes on a canonical thermal form
when expressed in terms of these eigenfunctions \thermal.
$$\eqalign{
G(r,\theta\vert r',\theta') &= {1\over 2 \pi}
K_0\left(m\sqrt{r^2+r'^2-2rr'\cos(\theta-\theta')}\right)\cr
&= \sum_{n = -\infty}^\infty \int_0^\infty {dE
\over 2E} e^{-E \vert \theta-\theta'+2\pi n \vert} \phi_E(r) \phi_E(r')\cr}
$$

Entropy of entanglement, relative to a division of space at $x=0$, is
defined as the entropy of the density matrix obtained by tracing out
the degrees of freedom located at $x<0$ from the vacuum density matrix
$\vert 0 >< 0 \vert$.  Set $\phi_0(x) = \phi_-(x) \theta(-x) +
\phi_+(x) \theta(x)$.  By expressing $\phi_-$ and $\phi_+$ in terms of
the Rindler eigenfunctions one finds that the vacuum wavefunctional is
an infinite product of harmonic oscillator propagators,
$$\eqalign{
\psi[\phi_-,\phi_+] &= \det{}^{1/2} G_D \,\, \exp - \half \int_0^\infty \,
  {E dE \over \sinh \pi E}  \Big[\left(\phi_+^2(E)+\phi_-^2(E)\right)
      \cosh \pi E \cr
 &\qquad\qquad\qquad\qquad\qquad\qquad\qquad\qquad\qquad
   - 2 \phi_+(E) \phi_-(E) \Big]\cr
 &= \prod_E <\phi_-(E) \vert e^{-\pi H_E} \vert \phi_+(E)>\cr}
$$
where $H_E$ is the quantum mechanical Hamiltonian for a harmonic
oscillator of frequency $E$.  The oscillator is evolved through $\pi$
in imaginary time, from an initial coordinate $\phi_+(E)$, to a final
coordinate $\phi_-(E)$.  A precise definition of the product over $E$
could be given by putting the system in a box to make the spectrum
discrete.

This shows that the entanglement density matrix is a thermal ensemble
with respect to the Rindler Hamiltonian at an inverse temperature
$\beta = 2 \pi$.  This is of course the Unruh effect \Unruh, that the
Minkowski vacuum state of a quantum field is a thermal state in
Rindler space.  It means that entropy of entanglement may be
calculated as thermal entropy in Rindler space.  Although the explicit
calculation we have performed is specific to a scalar field, the
formal proof that entropy of entanglement and Rindler thermal entropy
are the same is quite general \refs{\Israel,\thermal,\KabStr}.

To calculate the entropy of entanglement we are lead to construct
thermal ensembles with respect to the Rindler Hamiltonian at an
arbitrary temperature ($\beta \not= 2 \pi$).  The next step in making
contact with black hole entropy is to write the Rindler thermal
partition function as a determinant, which in turn becomes a
functional integral on `optical space'
\refs{\Ramond,\Barb,\Emp,\DeAlOh}.
$$\eqalign{
Z &= \det{}^{-1/2} \left( -\partial_\theta^2 + H_R^2 \right)\cr
  &= \int\limits_{\phi(\theta+\beta,r) = \phi(\theta,r)}
     \!\!\!\!\!\!{\cal D} \phi_{\rm optical}\,
     \exp - \int_0^\beta d\theta \int_0^\infty {dr \over r} \, \half
      \phi \left( -\partial_\theta^2 + H_R^2 \right)  \phi\cr}
$$
The eigenfunctions entering in the determinant are to be periodic in
$\theta$ with period $\beta$.  The action in the path integral is
exactly the action for a scalar field on a cone of deficit angle $2
\pi - \beta$.  The integration measure is different, however, since it
must be defined by
$$\eqalign{
&\int {\cal D} \phi_{\rm optical} \exp - \half \int d^2 x \sqrt{g_{\rm
optical}} \, \phi^2 = 1\cr
&ds^2_{\rm optical} = {1 \over r^2} \left(dr^2 +r^2 d\theta^2\right)
\qquad 0 \leq \theta \leq \beta \cr}
$$
This optical measure is what one would obtain by writing the action
for a scalar field on a cone in first order formalism, with
coordinates $\phi(x)$ and momenta $\pi(x)$, and adopting the canonical
measure ${\cal D}\pi {\cal D}\phi = \prod\limits_x {d \pi_x d\phi_x
\over 2 \pi}$.  Integrating out the momenta leaves the measure ${\cal
D} \phi_{\rm optical}$ for the integration over coordinates \DeAlOh.

Rather than use these optical/canonical measures, we would like to do
our calculations on a cone using a manifestly coordinate covariant
measure.
$$\eqalign{
&\int {\cal D} \phi_{\rm covariant} \exp - \half \int d^2 x
      \sqrt{g_{\rm cone}}\,\phi^2 = 1\cr
&ds^2_{\rm cone} = dr^2 + r^2 d\theta^2 \qquad 0 \leq \theta
\leq \beta \cr}
$$
Evidently there is a conflict between manifest covariance and the
canonical measure.  As the conical and optical metrics are related by
a conformal transformation, $ds^2_{\rm cone} = r^2 ds^2_{\rm
optical}$, with a conformal factor that is independent of $\theta$,
the corresponding integration measures differ by the exponential of a
term of ${\cal O}(\beta)$ (a Liouville action in the case of a
massless scalar field \DeAlOh).  This difference does not affect the
thermodynamics, which establishes that the entropy of entanglement of
a scalar field may be computed from its covariant partition function
on a cone.

We conclude this section with a calculation of the scalar partition
function \OnCones.  We work in arbitrary dimension on a space which is
a cone of deficit angle $2\pi - \beta$ times a transverse flat $(d-2)$
dimensional space with area $A_\perp$.  The free energy can be
written with an integral representation for the logarithm,
\eqn\ScalarBFexp{\eqalign{
\beta F &= \half \log \det (-\laplace + m^2)\cr
        &= - \half \int d^2x \sqrt{g} \int_{\epsilon^2}^\infty {ds \over s}
           e^{- s m^2} K(s,x,x)\, ,\cr}}
where the heat kernel $K(s,x,x)$ may be expressed as a single particle
path integral.
$$\eqalign{
K(s,x,x') &= <x \vert e^{-s(-\scriptlap)}\vert x'>\cr
 &= \int\limits_{x(0)=x' \atop x(s) = x}{\cal D} x(\tau) \,\,
     \exp - \int_0^s d\tau \, {1\over 4} g_{\mu\nu}
          \dot{x}^\mu \dot{x}^\nu \cr}
$$
To evaluate the heat kernel we diagonalize the Laplacian.
\eqn\ScalarEftns{\eqalign{
&\phi_{kl\kp}(r,\theta) = {e^{i \kp \cdot \xp} \over (2\pi)^{(d-2)/2}}
 (k/ \beta)^{1/2} \, e^{i 2 \pi l \theta / \beta}
  J_{\vert 2 \pi l / \beta\vert}(kr)\cr
&-\laplace \phi_{kl\kp} = \left(k^2+k_\perp^2\right) \phi_{kl\kp}
\qquad l \in \integer, \quad k \in \real^+, \quad \kp \in \real^{d-2}\cr
\noalign{\vskip 6pt}
&\int d^d x \sqrt{g} \, \phi_{kl\kp}^* \phi_{k'l'{\bf k}_\perp'}
 = \delta_{l\,l'}\,\delta(k-k')\,\delta^{(d-2)}(\kp - {\bf k}_\perp')\cr}}
We have chosen eigenfunctions which are regular at the origin,
corresponding to the Friedrichs extension of the Laplacian on a cone
\KayStu.  The heat kernel is given in terms of these eigenfunctions as
$$
K(s,x,x') = \sum_{l=-\infty}^\infty \int_0^\infty dk \int d^{d-2}k_\perp
  e^{-s \left(k^2+k_\perp^2\right)} \phi_{kl\kp}(x) \phi_{kl\kp}^*(x')\,.
$$
The sum and integral may be performed \refs{\Dowker,\DesJac},
with the result that at coincident points
\eqn\ScalarK{\eqalign{
K(s,x,x) & = {1 \over (4 \pi s)^{d/2}} - {1 \over 2 \beta} {1 \over
 (4 \pi s)^{d/2}}\int_{-\infty}^\infty dy \, e^{-{r^2 \over s}
 \cosh^2(y/2)} \cr
&\qquad\qquad\qquad \left(
 \cot {\pi \over \beta}(\pi + iy) + \cot {\pi \over \beta}(\pi - iy)
 \right)\,.\cr}}
The first term $1/(4 \pi s)^{d/2}$ in the heat kernel gives rise to a
divergence in the free energy, which can be absorbed in a
renormalization of the cosmological constant, and which does not
affect the entropy.  From the remainder of the heat kernel we find
\eqn\ScalarBFans{
 \beta F_{\rm scalar} = - {\pi^2 \over 3 \beta} A_\perp
 \left(1 - \left({\beta \over 2
 \pi}\right)^2 \right) \int_{\epsilon^2}^\infty {ds \over (4\pi s)^{d/2}}
 e^{-s m^2}\,.}
This is the result given in the Introduction.

\newsec{Vector Fields}

We now show that the entropy of entanglement of an Abelian gauge field
in two dimensions is identically zero, while the black hole entropy is
given by a non-zero surface term at the tip of the cone.

It is easiest to see that the entropy of entanglement vanishes in
Coulomb gauge.  In Cartesian coordinates on the plane, Coulomb
gauge corresponds to setting $A_\tau = 0$ and $\partial_x A_x = 0$.
There are no dynamical degrees of freedom in this gauge, so there can
be no correlations present across a boundary, and the entropy of
entanglement unambiguously vanishes.\foot{A possible constant
background electric field does not change this conclusion \BalCE.}

Defining the Rindler thermal entropy of a gauge field is more subtle.
Introduce polar coordinates on the plane.  It is tempting to adopt
Coulomb gauge, by setting $A_\theta = 0$ and $\partial_r A_r = 0$.  It
seems that there are no degrees of freedom in this gauge, so the
Rindler thermal entropy must vanish.  But Coulomb gauge breaks down at
the origin, where $A_\theta$ is ill-defined, so this argument is only
valid in the bulk: it shows that the bulk Rindler thermal entropy of a
two dimensional gauge field vanishes.

In order to include the origin in our treatment, we must use a
different gauge.  We now re-analyze the problem in covariant
gauge.  What we shall find is that, instead of unambiguously
vanishing, the Rindler thermal entropy is ill-defined, because the
Rindler Hamiltonian does not have normalizeable eigenstates.

In covariant gauge we diagonalize the quantum mechanical Rindler
Hamiltonian as follows.  The equations of motion in covariant gauge
$\nabla_\mu F^{\mu\nu} + {1 \over \xi} \nabla^\nu \nabla_\mu A^\mu =
0$ are solved by the ansatz
$$\eqalign{
A^L_\mu &= \partial_\mu \phi\cr
A^T_\mu &= \sqrt{g} \epsilon_{\mu \nu} \partial^\nu \phi\, ,\cr}
$$
where $\phi$ is a solution of the scalar equation of motion $\laplace
\phi = 0$.  We seek solutions of the form $A_\mu(r,\theta) =
e^{-E\theta} A_{E\,\mu}(r)$.  Given an eigenfunction $\phi_E(r)$ of
the scalar Rindler Hamiltonian \ScalarHR~we can construct a zero mode
of the Laplacian, $\phi_E(r,\theta) = e^{-E\theta} \phi_E(r)$, and in
turn a pair of eigenfunctions of the vector Rindler Hamiltonian.
$$\eqalign{
e^{-E\theta} A^L_{E\,\mu}(r) &= \partial_\mu \phi_E(r,\theta)\cr
e^{-E\theta} A^T_{E\,\mu}(r) &= \sqrt{g} \epsilon_{\mu \nu} \partial^\nu
\phi_E(r,\theta)\cr}
$$
These vector eigenfunctions are not acceptable, however.  As $r
\rightarrow 0$, they behave like $1/r$, and this makes them too
singular to be $\delta$-function normalizeable in the inner product
$$
<A \vert B> = \int_0^\infty {dr \over r} \, g^{\mu\nu} A_\mu^* B_\nu \,.
$$

We see that the quantum mechanical Rindler Hamiltonian is ill-defined
in covariant gauge, and that a thermal partition function cannot be
constructed, due to the bad behavior of the eigenfunctions near the
origin.  In this way a vector field escapes the formal proof that
entropy of entanglement is equal to black hole entropy \refs{\Israel,
\SussUg, \CalWil, \KabStr, \thermal}.

One can nevertheless define the partition function for a gauge field
on a cone using a path integral \Vass.  In the absence of a
Hamiltonian description of this path integral, the entropy obtained by
varying with respect to the deficit angle of the cone can and does
come out negative.  We begin in two dimensions and work in covariant
gauge.  The free energy may be expressed as a determinant.
$$\eqalign{
\beta F_{\rm gauge} &= \beta F_{\rm vector} + \beta F_{\rm ghosts}\cr
  &= \half \log \det \left(g^{\mu\nu}(-\laplace)
        -R^{\mu\nu} + \left(1-{1 \over \xi}\right)\nabla^\mu\nabla^\mu\right)
     - \log \det (-\laplace)\cr}
$$
To calculate the determinant we diagonalize the vector wave operator.
The explicit curvature term $R^{\mu\nu}$ in the wave operator is a
delta function at the tip of the cone; to treat this in a well defined
way we delete the point at the tip, and work on the space
$\real^2-\lbrace 0 \rbrace$.  Eigenfunctions of the vector wave
operator may then be generated from eigenfunctions of the scalar
Laplacian \ScalarEftns, using the same ansatz we used above
to solve the equations of motion.
\eqn\VectEftn{\eqalign{
&\nullcases{
 A^L_{kl} (r,\theta) = {1 \over k} \partial_\mu \phi_{kl}(r,\theta)
 & with eigenvalue ${1 \over \xi} k^2$ \cr
 A^T_{kl} (r,\theta) = {1 \over k} \sqrt{g} \epsilon_{\mu \nu} \partial^\nu
 \phi_{kl} (r,\theta) & with eigenvalue $k^2$ \cr}
\quad k \in \real^+, \,\, l \in \integer\cr
\noalign{\vskip 6pt}
&\int d^2x \sqrt{g} g^{\mu\nu} A_{kl\,\mu}^{L\,*} A_{k'l'\,\nu}^L
 = \int d^2x \sqrt{g} g^{\mu\nu} A_{kl\,\mu}^{T\,*} A_{k'l'\,\nu}^T
 = \delta_{ll'} \delta(k-k')\cr}}
The vector determinant can be expressed in terms of these eigenfunctions
via a heat kernel.  It is necessary to introduce both
ultraviolet and infrared cutoffs in $d=2$.
\eqn\VectBF{\eqalign{
\beta F_{\rm vector} &= - \half \int d^2x \sqrt{g} \bigg(
\int_{\epsilon_L^2}^\infty
  {ds \over s} e^{-s m_L^2} g_{\mu\nu} K^{\mu\nu}_L(s,x,x)\cr
&\qquad\qquad\qquad + \int_{\epsilon_T^2}^\infty
  {ds \over s} e^{-s m_T^2} g_{\mu\nu} K^{\mu\nu}_T(s,x,x)\bigg)\cr
K^{\mu\nu}_L(s,x,x') &= \sum_{l=-\infty}^\infty \int_0^\infty dk
e^{-s k^2/\xi} A^\mu_L(x) A^{\nu\,*}_L(x')\cr
K^{\mu\nu}_T(s,x,x') &= \sum_{l=-\infty}^\infty \int_0^\infty dk
e^{-s k^2} A^\mu_T(x) A^{\nu\,*}_T(x')\cr}}
We fix the remaining ambiguity in these expressions by imposing some
physical requirements.

First, the vector wave operator must be self-adjoint.  When a vector
field is expressed in terms of a scalar field using the ansatz
\VectEftn, this becomes the requirement that the scalar field be
non-singular at the tip of the cone.  So in constructing the vector
determinant, we must use the non-singular scalar eigenfunctions
\ScalarEftns.

Second, we must impose BRST invariance,
$$
  \delta_\chi \,A_\mu = i \chi \partial_\mu \eta_2 \qquad
  \delta_\chi \,\eta_1 = {1 \over \xi} \chi \nabla_\mu A^\mu \qquad
  \delta_\chi \,\eta_2 = 0
$$
where $\eta_1$, $\eta_2$ are real scalar ghosts and $\chi$ is a
Grassmann parameter.  This requires that the ghost determinant be
constructed from the same scalar eigenfunctions that enter in the
vector determinant.  Note that BRST invariance plus self-adjointness
implies that $\beta F_{\rm ghost} = -2 \beta F_{\rm scalar}$, with the
scalar determinant given in \ScalarBFans.

Third, we'd like the partition function to be independent of the
gauge fixing parameter $\xi$.  The partition function is gauge
invariant; it will be independent of $\xi$ provided we use a gauge
invariant cutoff.  Suppose we regulate the ghost determinant as in
\ScalarBFexp, with an infrared cutoff $m$ and an ultraviolet cutoff
$\epsilon$.  If we set the cutoffs on the vector determinant
\VectBF~according to
$$\eqalign{
&\epsilon_L^2 = \xi \epsilon^2, \qquad m_L^2 = m^2/\xi\cr
&\epsilon_T^2 = \epsilon^2, \qquad m_T^2 = m^2\cr}
$$
then $\xi$ will drop out of the free energy, upon rescaling the proper
time $s \rightarrow \xi s$.  No matter what choice of covariant gauge
one makes initially, with this choice of cutoffs one ends up in
Feynman gauge, with $\xi=1$.

In fact this choice of cutoffs is not arbitrary.  We have constructed
regulated determinants, defined by
$$
\log \det L \equiv {\rm Tr} \left( e^{-\epsilon^2 L} \log \epsilon^2
  (L+m^2)\right)\,.
$$
A BRST transformation relates a ghost mode with eigenvalue $k^2$ to a
longitudinal mode with a {\it different} eigenvalue $k^2/\xi$.  To
respect BRST invariance the regulator must modify these two
eigenvalues in the same way, which fixes the above relationship
between the ghost and longitudinal cutoffs.

We continue with the calculation of the partition function, by
expressing the spin-traced vector heat kernel at coincident points in
terms of the eigenfunctions \VectEftn.  We set $\xi = 1$.
$$
g_{\mu\nu} K^{\mu\nu}_{\rm vector}(s,x,x) = 2 \sum_{l=-\infty}^\infty
\int_0^\infty {dk \over \beta k} e^{-s k^2}
\left(\big(\partial_r J_{\vert 2 \pi l / \beta \vert}(kr)\big)^2
+ \left({2 \pi l \over \beta r} J_{\vert 2 \pi l / \beta \vert}(kr)
\right)^2\right)
$$
This is equal to the scalar heat kernel plus a total derivative.
$$\eqalign{
g_{\mu\nu} K^{\mu\nu}_{\rm vector}(s,x,x) &= 2 K_{\rm scalar}(s,x,x)
+ \sum_{l=-\infty}^\infty
\int_0^\infty {dk \over \beta k} e^{-s k^2}
{1 \over r} \partial_r r \partial_r
\big(J_{\vert 2 \pi l / \beta \vert}(kr)\big)^2\cr
&= 2 K_{\rm scalar}(s,x,x)
+ {1 \over r} \partial_r \int_s^\infty ds' r \partial_r
K_{\rm scalar}(s',x,x)\cr
&= 2 K_{\rm scalar}(s,x,x) + {2 \over r} \partial_r
s K_{\rm scalar}(s,x,x) \cr}
$$
The last equality may be checked from the explicit expression for
$K_{\rm scalar}$ given in \ScalarK.  It is a consequence of
dimensional analysis; on dimensional grounds the scalar heat kernel
takes the form ${1 \over s'} f(r^2/s')$, so we can replace $r
\partial_r$ with $- 2 \partial_{s'} s'$, making the $s'$ integral
trivial.  The free energy becomes
$$\eqalign{
\beta F_{\rm vector} &=
- \int d^2x \sqrt{g} \int_{\epsilon^2}^\infty {ds \over s}
  e^{-s m^2} K_{\rm scalar}(s,x,x)\cr
& \quad + \beta \int_{\epsilon^2}^\infty ds  e^{-s m^2}
       \left(K_{\rm scalar}(s,0,0) - {1 \over 4 \pi s}\right) \,.\cr }
$$
The term $1/4 \pi s$ is a surface term from $r=\infty$, which will
cancel the surface term from $r=0$ if $\beta = 2 \pi$.  The bulk
contribution from the vector cancels against the ghosts in $d=2$.  The
scalar heat kernel was explicitly evaluated in \ScalarK; representing
it as a single particle path integral leads to the result given in the
Introduction, that the free energy for a gauge field in two dimensions
is given solely by a surface term, which is an integral over particle
paths beginning and ending on the horizon.
$$\eqalign{
\beta F_{\rm gauge} &=  \beta F_{\rm vector} + \beta F_{\rm ghosts}\cr
&= \beta \int_{\epsilon^2}^\infty ds \left[
  \int\limits_{\,\,x(0) = x(s) = 0}\!\!
       {\cal D}x(\tau)\,\, \exp - \int_0^s d\tau ({1\over 4} g_{\mu\nu}
       \dot{x}^\mu \dot{x}^\nu + m^2) - {e^{-s m^2}
\over 4 \pi s} \right] \cr
&= (2\pi - \beta)
          \int_{\epsilon^2}^\infty {ds \over 4 \pi s} e^{-s m^2}\cr}
$$

It is straightforward to extend this result to higher dimensions.  We
will only discuss Feynman gauge $\xi=1$, since the gauge fixing
parameter drops out of the free energy with a suitable choice of
cutoffs, just as it did in two dimensions.  It is convenient to
introduce indices in the $r$--$\theta$ plane, $\alpha,\,\beta =
r,\,\theta$, and indices in the transverse flat directions, $i,\,j =
1,\ldots,d-2$.  The vector field $A_\mu$ decomposes into a two
dimensional vector $A_\alpha$, and a collection of $d-2$ scalar fields
$A_i$.  The vector wave operator is block diagonal in Feynman gauge.
$$\eqalign{
(-\laplace A)_\alpha &= (-\nabla_\beta\nabla^\beta - \partial_j
   \partial^j) A_\alpha\cr
(-\laplace A)_i &= (-\nabla_\beta\nabla^\beta - \partial_j \partial^j)
A_i\cr}
$$
Note the distinction: in the first line, the covariant derivative acts
on a vector, while in the second it acts on a scalar.  The
eigenfunctions of the vector wave operator can be expressed in terms
of the scalar eigenfunctions \ScalarEftns.
$$\eqalign{
&A_\alpha^L = {1 \over k} \partial_\alpha \phi_{kl\kp},\;\;
              A_i^L = 0\cr
&A_\alpha^T = {1 \over k} \sqrt{g} \epsilon_{\alpha\beta}
              \partial^\beta \phi_{kl\kp},\;\; A_i^T = 0\cr
&A_\alpha^{(i)} = 0,\;\; A_i^{(i)} =
              \delta_i^{(i)}\phi_{kl\kp}\qquad (i)=1,\ldots,d-2\cr}
$$
This leads to an expression for the spin-traced vector heat kernel in
terms of the scalar heat kernel.
$$\eqalign{
g_{\mu\nu} K^{\mu\nu}_{\rm vector}(s,x,x) &= (d-2) K_{\rm scalar}(s,x,x)\cr
&\qquad + 2 \sum_{l=-\infty}^\infty \int_0^\infty dk \int d^{d-2}k_\perp
e^{-s\left(k^2+k_\perp^2\right)} {1 \over k^2} g^{\alpha\beta}
\partial_\alpha \phi_{kl\kp} \partial_\beta\phi_{kl\kp}^*\cr
&= d \, K_{\rm scalar}(s,x,x)\cr
&\qquad + {1 \over r} \partial_r \sum_{l=-\infty}^\infty
\int_0^\infty dk \int d^{d-2}k_\perp e^{-s\left(k^2+k_\perp^2\right)}
{1 \over k^2} r \partial_r \vert\phi_{kl\kp}\vert^2\cr
&= d \, K_{\rm scalar}(s,x,x) + {2 \over r} \partial_r s
K_{\rm scalar}(s,x,x)\cr}
$$
The ghosts cancel the bulk free energy of two scalar fields, and one
obtains the result given in the Introduction.
$$\eqalign{
\beta F_{\rm gauge} &= (d-2) \left(-\half\right) \int_{\epsilon^2}^\infty
    {ds \over s}
    \int\limits_{\hbox{\eightrm closed loops}}
     \!\!{\cal D}x(\tau)\,\, \exp - \int_0^s d\tau \,{1\over 4} g_{\mu\nu}
       \dot{x}^\mu \dot{x}^\nu\cr
&\qquad + \beta\int_{\epsilon^2}^\infty ds \!\!
 \int\limits_{\,\,\,\,x(0) = x(s) \in\, {\rm horizon}}
 \!\!\!\!\!\!\!\!{\cal D}x(\tau)\,\, \exp - \int_0^s d\tau \,
   {1\over 4} g_{\mu\nu} \dot{x}^\mu \dot{x}^\nu \cr
&\qquad - \beta A_\perp \int_{\epsilon^2}^\infty {ds \over
   (4 \pi s)^{d/2}} \cr
&= (d-2) \beta F_{\rm scalar} + A_\perp (2\pi - \beta)
          \int_{\epsilon^2}^\infty {ds \over (4 \pi s)^{d/2}}\cr}
$$
A few remarks on this result:
\item{(i)}  The surface terms cancel at $\beta = 2 \pi$, which is
fortunate, because there should not be a contact interaction
with the horizon when the curvature there vanishes.
\item{(ii)}  There are different measures for the integral over proper
time in the surface free energy compared to the bulk.  The additional
$1/s$ is present in the bulk because the group of global isometries of
a circle, $\tau \rightarrow \tau + const.$, must be gauge fixed when
performing a path integral over closed particle paths \Poly.

As we shall see in the next section, no surface term arises for spinor
fields.  Note that the vector surface term arises from the behavior of
the fields in the two dimensions which contain the conical
singularity.  The reason gauge fields differ from spinors and scalars
is that they are not conformally invariant in $d=2$.  All fields
become effectively massless close to the tip of the cone, where the
proper temperature goes to infinity.  Scalar and spinor fields are
conformally invariant in this limit, and one may use techniques from
conformal field theory to obtain their entropy \HoLaWi.  A conformal
mapping from the cone to a cylinder shows that, for any conformal
field theory, the leading divergence of the entropy is insensitive to
the boundary conditions imposed at the tip of the cone.  This argument
does not apply to gauge fields, where a singular boundary term arises.
For example, cutting out a small disc around the tip of the cone and
imposing boundary conditions on the resulting edge would dramatically
change the vector partition function by eliminating the surface term.

\newsec{Spinor Fields}

Having treated spins zero and one, it is natural to inquire as to what
happens for spin one-half.  We now show that no surface term arises,
and that the contribution of a fermion to the entropy of a black hole
is equal to its entropy of entanglement as well as its Rindler thermal
entropy.  We work in two dimensions.

We first ask whether there is a well defined Rindler Hamiltonian for
fermions.  Introduce the standard zweibein and spin connection on the cone.
\eqn\ConeZwei{\eqalign{
ds^2 &= dr^2 + r^2 d\theta^2 \qquad 0 \leq \theta \leq \beta\cr
e^1 &= \cos \left({2 \pi \over \beta} \theta\right) dr - r \sin
\left({2 \pi \over \beta} \theta \right) d\theta\cr
e^2 &= \sin \left( {2 \pi  \over \beta} \theta \right) dr + r \cos \left(
{2 \pi \over \beta} \theta \right)  d\theta\cr
\omega &= \left(1 - {2 \pi \over \beta}\right) d \theta\cr}}
The spin connection has been chosen so that parallel transport around
the tip of the cone generates a rotation through the deficit angle
$2\pi - \beta$.  At $\beta = 2 \pi$ this zweibein reduces to the
usual Cartesian basis of flat space, and the spin connection
vanishes.  An inequivalent spin structure on the cone is defined by
\eqn\OtherZwei{
e^1 = dr,\quad e^2 = r d\theta,\quad \omega = d\theta\,.}
The two spin structures differ by a topologically non-trivial Lorentz
rotation, which winds once around the Lorentz group as the origin is
encircled.  This will be important below.

The Euclidean Dirac equation $(i \slash{\nabla} - m)\psi = 0$ on a
cone may be presented in the form\foot{Conventions: $\gamma^1 = i
\sigma^1$, $\gamma^2 = i \sigma^2$, $\gamma^5 = i \gamma^1 \gamma^2 =
\sigma^3$, $\bar\psi = \psi^\dagger \gamma^2$.  $\Sigma_{ab} = {1 \over 4}
[\gamma^a,\gamma^b]$, $\nabla_\mu = \partial_\mu + \half
\omega_\mu^{ab} \Sigma_{ab} = \partial_\mu - {i \over 2} \omega_\mu
\gamma^5$.}
$$
\exp(-i\pi\theta \gamma^5/\beta) \, \gamma^2
(\partial_\theta + H_R )\exp(i\pi\theta\gamma^5/\beta) \psi = 0
$$
where the Rindler Hamiltonian is explicitly given by
$$
H_R = \left(\matrix{&-i(r \partial_r + \half) & -imr \cr & imr &
i(r\partial_r+\half) \cr}\right) \,.
$$
The rotation operators $e^{i \pi \theta \gamma^5 / \beta}$ are present
because the zweibein \ConeZwei~depends on $\theta$.  The rotation
operators make up for this, by rotating $\psi$ to the
$\theta$-independent frame \OtherZwei, where the rotational invariance
of the cone is manifest, and rotations are generated by a Rindler
Hamiltonian that is independent of $\theta$.  $H_R$ is self-adjoint in
the inner product $<\psi_1\vert\psi_2> = \int_0^\infty dr
\psi_1^\dagger(r)\psi_2(r)$, and has a complete set of eigenfunctions.
$$\eqalign{
&\psi_E(r) = {1 \over \pi} \sqrt{m \cosh \pi E}
\left(\matrix{&K_{iE-1/2}(mr)\cr &K_{iE+1/2}(mr)\cr}\right)\cr
&H_R \psi_E = E \psi_E \qquad -\infty < E < \infty\cr
&\int_0^\infty dr \psi_E^\dagger \psi_{E'} = \delta(E-E')\cr}
$$
We see that there is no difficulty in constructing a Rindler
Hamiltonian for fermions.  The formal argument showing that entropy of
entanglement is identical to Rindler thermal entropy applies to
fermions \refs{\Israel, \SussUg, \CalWil, \KabStr, \thermal}.  To
construct the entropy of entanglement for fermions directly from its
definition one must introduce a functional state space for fermions
\FermiSchr.  The entropy of entanglement for fermions has been
calculated directly from its definition by Larsen and Wilczek \LarWil.

To make contact with black hole entropy, we write the Rindler thermal
partition function as a determinant,
$$
\beta F_{\rm Dirac} = - \half \log \det (-\partial_\theta^2 + H_R^2)\,,
$$
where the determinant is over functions which are anti-periodic in
$\theta$.  This determinant may be represented as a functional
integral over anti-periodic Grassmann fields.
$$\eqalign{
\beta F_{\rm Dirac} &= - \log \det i\left(\partial_\theta + H_R\right)\cr
&= - \log \int {\cal D} \psi {\cal D} \psi^\dagger \exp \int_0^\beta d\theta
\int_0^\infty dr\,\psi^\dagger \, i \left(\partial_\theta + H_R\right)\psi\cr
&= - \log \int {\cal D}\psi {\cal D} \bar\psi \exp - \int d^2x
\sqrt{g_{\rm cone}}\,
\bar\psi \, {i \over r} \gamma^2 \left(\partial_\theta + H_R\right)\psi\cr
&= - \log \int {\cal D}\psi {\cal D} \bar\psi \exp - \int d^2x
\sqrt{g_{\rm cone}}\, \bar\psi \, (i \slash{\nabla} - m) \psi\cr}
$$
One gets the classical action for a fermion on a cone, but {\it with
the other choice of spin structure}.  That is, in the last line
$\slash{\nabla}$ is constructed from \OtherZwei, which differs from
the standard spin connection \ConeZwei~by a Lorentz rotation which
varies from zero to $2\pi$ as the origin is encircled.  One may revert
to the standard spin connection on a cone, at the price of changing
the fermion boundary conditions from anti-periodic to periodic.  So in
the end we have an expression for entropy of entanglement in terms
of a functional integral over fermions which are periodic on a cone.

There is also the question of the appropriate measure in the
functional integral.  As in the scalar case, the measure that
reproduces the Rindler thermal partition function is the canonical
measure on a cone, which differs from the covariant measure by a term
of ${\cal O}(\beta)$ in the free energy.  This difference does not
affect the thermodynamics, so we will neglect it, and proceed to
calculate the fermion free energy using the covariant measure.

The free energy can be expressed in terms of a heat kernel.
$$\eqalign{
\beta F &= - \log \det (i \slash{\nabla} - m)\cr
        &= - \half \log \det (\slash{\nabla}^2 + m^2)\cr
        &= \half \int d^2x \sqrt{g} \int_{\epsilon^2}^\infty {ds \over s}
           e^{-s m^2} {\rm Tr}\, K_{\rm Dirac}(s,x,x)\cr}$$
The heat kernel is constructed from eigenfunctions of the Dirac
operator.  For $\beta < 2 \pi$ the Dirac operator has a complete set of
non-singular eigenfunctions \GerJac.
$$\eqalign{
&\psi_{k l} = \sqrt{k \over 2 \beta} \left(\matrix{
& e^{i 2 \pi l \theta / \beta} J_{\vert \nu \vert}(k r) \cr
& -i \, {\rm sign}(\nu) e^{i 2 \pi (l+1) \theta / \beta} J_{\vert \nu \vert +
 {\rm sign}(\nu)}(k r) \cr}\right)\cr
&\quad \nu \equiv {2 \pi \over \beta} (l+\half) - \half, \qquad k \in
\real^+,\quad l \in \integer\cr
&\slash{\nabla} \psi_{k l} = k \psi_{k l}\qquad
\slash{\nabla} \gamma^5 \psi_{k l} = - k \gamma^5 \psi_{k l}\cr
&\int d^2x \sqrt{g} \, \psi_{kl}^\dagger \psi_{k'l'} = \delta_{ll'}
\delta(k-k')\cr}
$$
For $\beta > 2 \pi$ these eigenfunctions develop a singularity at the
origin.  In fact there are no non-singular eigenfunctions of the Dirac
operator when $\beta > 2 \pi$, as using the other set of Bessel
functions $Y_\nu(z)$ also leads to singular eigenfunctions.  This
necessitates a non-trivial self-adjoint extension of the Dirac
operator \GerJac.  The self-adjoint extension parameter must be chosen
so as to make the free energy analytic across $\beta=2\pi$.  A simpler
procedure, which we will adopt, is to calculate for $\beta < 2\pi$ and
define the free energy for $\beta>2\pi$ by analytic continuation.
The heat kernel may be evaluated using the same techniques as in the
scalar case \refs{\Dowker,\DesJac}.  At coincident points the spin
traced heat kernel is given by
$$\eqalign{
{\rm Tr}\, K_{\rm Dirac}(s,x,x)  &= {\rm Tr}\,\sum_{l=-\infty}^\infty
\int_0^\infty dk e^{-sk^2}
\left(\psi_{k l}(x)\psi^\dagger_{k l}(x) +
\gamma^5 \psi_{k l}(x)\psi^\dagger_{k l}(x)\gamma^5\right)\cr
&= {1 \over 2 \pi s} + {1 \over \beta}
{1 \over 2 \pi s} \int_{-\infty}^\infty dy e^{-{r^2 \over s} \cosh^2(y/2)}
{i \sinh {y \over 2} \over \sin{\pi \over \beta}(\pi + i y)}\,.\cr}
$$
Substituting this into the expression for the free energy, we find the
same free energy as a scalar field in two dimensions, aside from a
difference in the cosmological constant which we neglect.
$$\beta F_{\rm Dirac} =  - {\pi^2 \over 3 \beta} \left(1 -
\left({\beta \over 2 \pi}\right)^2\right)
\int_{\epsilon^2}^\infty {ds \over 4 \pi s} e^{-s m^2}
$$
This result was given in the Introduction; it leads to the same
entropy at $\beta=2\pi$ as was calculated from the renormalization of
Newton's constant in $d=2$.

\newsec{Conclusions}

In exploring the relationship between black hole entropy and entropy
of entanglement, we have seen that, for scalar and spinor fields, the
two are identical, while for gauge fields, they differ by a contact
interaction with the horizon which appears in the black hole entropy
but not in the entropy of entanglement.  The contact interaction makes
a negative contribution to the entropy of a black hole, and is
responsible for the non-renormalization of Newton's constant in
certain supersymmetric theories.  It does not have a thermal or state
counting interpretation within quantum field theory.  It is the field
theoretic residue of the interaction proposed by Susskind and Uglum
which couples a closed string to an open string stranded on the
horizon.

The classical entropy of a black hole also arises from a contact term
localized on the horizon \refs{\BTZ,\CT}.  A state counting
interpretation of these contact terms is only possible if quantum
gravity introduces a degree of non-locality, which resolves the
point-like contact terms into extended interactions for which a notion
of state counting can exist.  It seems plausible that such an effect
does occur, as one would expect quantum gravity to de-localize the
horizon by at least a Planck length.  A concrete proposal has been put
forth by Susskind and Uglum, in which the extended nature of
fundamental strings provides a resolution of the contact interactions
and leads to a state counting interpretation of the classical black
hole entropy \SussUg.

We are led to investigate these issues in theories which are non-local
at short distances.  One way to obtain such behavior within ordinary
quantum field theory is to study the behavior of a composite field,
constructed in a non-local way from elementary fields.  Such
constructions arise naturally in the low energy description of certain
large $N$ field theories \KSS.

\bigbreak
\bigskip
\bigskip
\centerline{\bf Acknowledgments}
\nobreak
I am grateful to my collaborators Steve Shenker and Matt Strassler for
their many suggestions, and to Roberto Emparan and Leonard Susskind
for valuable discussions.  This work was supported by the D.O.E.~under
contract \#DE--FG05--90ER40559.  Part of this work was completed at
the Aspen Institute for Physics.

\listrefs
\bye